# Depolarizing-Field Effect in Strained Nanoscale Ferroelectric Capacitors and Tunnel Junctions

N. A. Pertsev and H. Kohlstedt

*Institut für Festkörperforschung, Forschungszentrum Jülich, D-52425 Jülich, Germany*

The influence of depolarizing field on the magnitude and stability of a uniform polarization in ferroelectric capacitors and tunnel junctions is studied using a nonlinear thermodynamic theory. It is predicted that, in heterostructures involving strained epitaxial films and metal electrodes, the homogeneous polarization state may remain stable against transformations into the paraelectric phase and into polydomain states down to the nanoscale. This result supports the possibility of depolarizing-field-related resistive switching in ferroelectric tunnel junctions with dissimilar electrodes. The resistance on/off ratio in such junctions is shown to be governed by the difference between the reciprocal capacitances of screening space charges in the electrodes.

## I. INTRODUCTION

Persistence of ferroelectricity in ultrathin films is an issue of high fundamental and practical interest. In particular, the stability of ferroelectric states with a nonzero net polarization represents a matter of primary importance. Indeed, the presence of considerable remnant out-of-plane polarization is necessary for the memory applications of ferroelectric films in the form of capacitors [1] and ferroelectric tunnel junctions (FTJs) [2-4].

Dependence of the ferroelectric polarization **P** on the thickness of a thin film sandwiched between two electrodes may result from both long-range and short-range interactions. The most widely discussed cause of this size effect is the existence of a depolarizing field appearing when the polarization charges $\rho = -\,\mathrm{div}\,\mathbf{P}$ at the film surfaces are not perfectly compensated for by free charge carriers [5-11]. The phenomenological theory predicted long time ago that this internal electric field differs from zero even in thin films covered by metal electrodes [6,8]. The depolarizing-field effect may lead to the instability of the out-of-plane polarization in ultrathin films resulting in complete disappearance of ferroelectric phase below some critical thickness [5,7]. This prediction of the mean-field theory was supported recently by first-principles calculations [10,11]. The depolarizing-field effect on polarization was invoked to explain gradual reduction of tetragonality measured in ultrathin $PbTiO_3$ films [12] and the time-dependent polarization relaxation observed in $SrRuO_3/BaTiO_3/SrRuO_3$ capacitors [13]. In addition to the long-range depolarizing field, the polarization in a thin film may be affected by the intrinsic surface effect associated with dipole-dipole interactions [14] and by short-range interactions between atomic layers adjacent to the ferroelectric-electrode interface [11].

In epitaxial films, which are most suited for the fabrication of ferroelectric capacitors and tunnel junctions, the polarization state is also strongly influenced by lattice strains induced by the mechanical film/substrate interaction [15,16]. Fortunately, the out-of-plane polarization is enhanced in epitaxial films of perovskite ferroelectrics grown on "compressive" substrates [17]. In this case the electrostrictive coupling between the polarization and compressive in-plane strains makes the out-of-plane polarization state more stable than in free standing films [18]. Accordingly, in epitaxial films grown on compressive substrates the depolarizing-field effect competes with the lattice-strain one. This competition must be taken into account in the studies of strained ferroelectric capacitors and FTJs.

The aim of this paper is to analyze the influence of depolarizing field on the physical properties of strained ferroelectric films sandwiched between metallic electrodes. In contrast to some recent studies [4,13], we use a nonlinear thermodynamic theory to calculate the depolarizing field and the film polarization in a metal-ferroelectric-metal (MFM) heterostructure self-consistently. The results of our calculations indicate that the nanoscale ferroelectric capacitors and FTJs can be employed for memory applications.

We focus on thin films of perovskite ferroelectrics epitaxially grown on a thick cubic substrate inducing compressive in-plane lattice strains in the film. At thicknesses outside the nanoscale range, such films stabilize in the tetragonal *c* phase with the polarization orthogonal to the film surfaces [15,16]. Since we are interested in the memory applications, the *c* phase is taken to be in a single-domain state. (This is an appropriate model because an electric field larger than the coercive field is applied to the film in a memory cell so that the 180º domains are mostly removed.) Neglecting for clarity the surface effects on polarization [11,14] and a weak conductivity of perovskite ferroelectrics at low temperatures, we can further assume the film to be homogeneously polarized. For such ferroelectric film, we first calculate the equilibrium polarization as a function of the film thickness and then determine the thickness range, where the single-domain state is stable with respect to the transformation into a polydomain one. It should be noted that in some part of this range the uniformly polarized state may appear to be energetically less favorable than the 180º domain pattern [19,20]. Such metastability, however, is not detrimental for memory applications, provided that the potential barrier between the aforementioned states is high enough to prevent fast relaxation of the remanent polarization at a low operating temperature.

## II. THICKNESS DEPENDENCE OF SPONTANEOUS POLARIZATION

To calculate the equilibrium polarization, we use a nonlinear equation of state of a strained ferroelectric film.

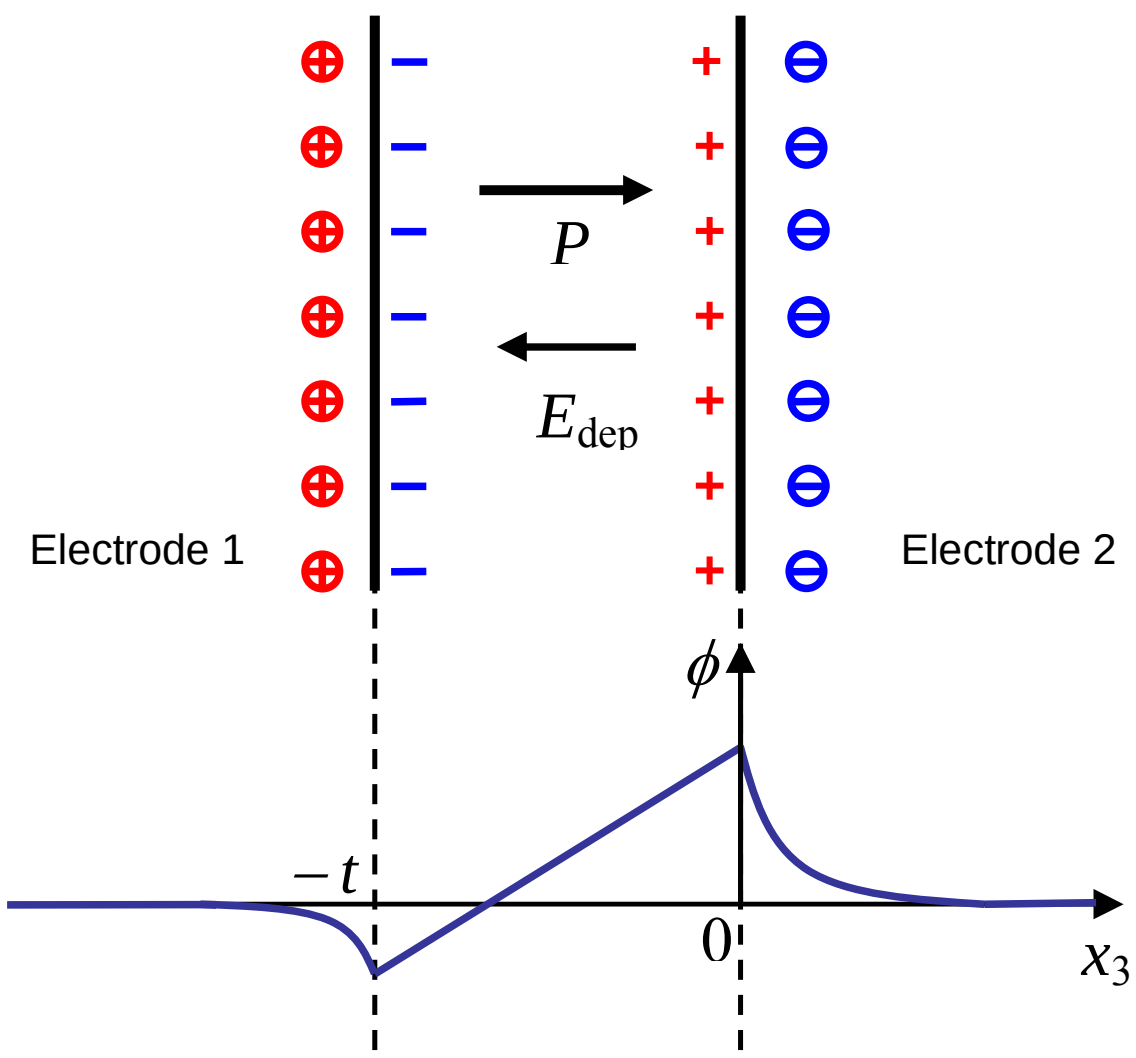

FIG. 1. Screening of polarization charges in a metal-ferroelectric-metal heterostructure. Distribution of the electrostatic potential $\phi$ is shown schematically for the case of dissimilar electrodes kept at a bias voltage compensating for the difference of their work functions.

Such equation can be derived by differentiating the film thermodynamic potential [15,16] or the Helmholtz free energy [17] written in terms of polarization components $P_i$ ($i$ = 1,2,3). For the *c* phase ($P_1 = P_2 = 0$, $P_3 \neq 0$) stable at negative misfit strains $S_m$, the relevant equation reduces to

$$2a_3^* P_3 + 4a_{33}^* P_3^3 + 6a_{111} P_3^5 + 8a_{1111} P_3^7 + ... = E_3, \quad (1)$$

where $a_3^* = a_1 + 2S_m(q_{11}c_{12} - q_{12}c_{11})/c_{11}$, $a_{33}^* = a_{11} - q_{11}^2/(2c_{11})$, $a_1$, $a_{ij}$, $a_{ijk}$, and $a_{ijkl}$ are the dielectric stiffness coefficients at constant strain, $q_{ln}$ are the relevant electrostrictive constants, and $c_{ln}$ are the film elastic stiffnesses at constant polarization. The field $E_3$ inside the film can be calculated from the total voltage drop across the MFM heterostructure and the continuity condition of the electric displacement $\mathbf{D} = \varepsilon_0 \mathbf{E} + \mathbf{P}$ at the ferroelectric-metal interfaces ($\varepsilon_0$ is the permittivity of the vacuum) [6,7]. This procedure was employed in many papers, but the linear approximation was frequently used for the field dependence of the ferroelectric polarization [3,4,8,13]. In our formulation, the screening of polarization charges results from changes of the electron density inside the electrodes. The electric field $E_m$ in a metal electrode can be evaluated in the Thomas-Fermi approximation [21] corrected for the dielectric response of metallic lattice [22]. For a thick electrode occupying the space $x_3 \geq 0$ (see Fig. 1), the calculation gives [8]

$$E_m(x_3) = E_m(x_3 = 0)\exp(-x_3 / l_s)\,. \tag{2}$$

Here $l_s$ is the screening length defined by the relation

$$l_s^2 = \left(\frac{3}{8\pi}\right)^{2/3} \frac{h^2}{3} \frac{\varepsilon_0 \varepsilon_m}{e^2 m^*} \frac{1}{n_0^{1/3}}\,, \tag{3}$$

where $n_0$ is the mean electron density at zero electric field, $\varepsilon_m$ is the relative permittivity of the ionic lattice in the metal, $e$ is the electronic charge, $m^*$ is the electron effective mass in the electrode, and $h$ is the Planck constant. The field $E_m$ at the boundary $x_3 = 0$ between the electrode and ferroelectric can be found from the electric displacement $D_m = \varepsilon_0\ \varepsilon_m\ E_m$ in the metal, taking into account that $D_m(x_3 = 0)$ is defined by the total screening charge $\rho_m$ in the electrode (per unit area). The integration of Eq. (2) then enables us to calculate the voltage change across the electrode to be $\Delta\phi_m = \rho_m l_s /(\varepsilon_0 \varepsilon_m)$. Introducing the capacitance $c_m$ per unit area of the screening space charge [23,24] as $c_m = \varepsilon_0\ \varepsilon_m / l_s$, we can write this relation in the compact form $\Delta\phi_m = \rho_m / c_m$.

The algebraic sum of the voltage drop $\Delta\phi_f = -E_3 t$ in a ferroelectric film of thickness $t$ and the voltage changes $\Delta\phi_{m1} = -\rho_{m1} / c_{m1}$ and $\Delta\phi_{m2} = \rho_{m2} / c_{m2}$ in the left and right electrodes is governed the applied voltage difference $V_a = V_2 - V_1$. The continuity of electrostatic potential yields

$$V_a = -\frac{\rho_{m1}}{c_{m1}} - E_3 t + \frac{\rho_{m2}}{c_{m2}} + \frac{W_2 - W_1}{e}\,, \tag{4}$$

where the last term accounts for the contact potential difference that appears when electrodes have different work functions $W_1$ and $W_2$ [25]. For the screening charges $\rho_{m1}$ and $\rho_{m2}$, the boundary conditions at the two ferroelectric/electrode interfaces give $\rho_{m1} = \varepsilon_0 E_3 + P_3$ and $\rho_{m2} = -(\varepsilon_0 E_3 + P_3)$. Substituting these relations into Eq. (4), we finally obtain the formula

$$E_3 = -\frac{P_3}{\varepsilon_0 + c_i t} - \frac{c_i}{(\varepsilon_0 + c_i t)}\left(V_a - \frac{W_2 - W_1}{e}\right) \tag{5}$$

for the total field inside the film [26]. The first term in Eq. (5), which is proportional to the film out-of-plane polarization, represents the depolarizing field $E_{dep}$. It can be seen that the only parameter of the electrodes that directly affects $E_{dep}$ is the *total interfacial capacitance* $c_i = (c_{m1}^{-1} + c_{m2}^{-1})^{-1}$ resulting from the presence of two capacitances $c_{m1}$ and $c_{m2}$ in series. Existence of an interfacial capacitance $c_i$ also reduces the second term in Eq. (5), which describes the external field acting on a ferroelectric film. However, this reduction is expected to be weak since usually $c_i t >> \varepsilon_0$.

The substitution of Eq. (5) into Eq. (1) shows that the depolarizing field formally renormalizes the coefficient $a_3^*$ of the lowest-order polarization term. This renormalization, which can be described by the introduction of the coefficient

$$a_3^{**} = a_3^* + \frac{1}{2(\varepsilon_0 + c_i t)} = a_1 + 2\left(\frac{q_{11} c_{12}}{c_{11}} - q_{12}\right) S_m + \frac{1}{2(\varepsilon_0 + c_i t)} \tag{6}$$

instead of $a_3^*$, should affect all physical properties of a single-domain ferroelectric film. However, the difference between the coefficients $a_3^{**}$ and $a_1$ may be reduced and even compensated by the strain effect. Indeed, the second term on the right-hand side of Eq. (6) is negative for films of perovskite ferroelectrics grown on an appropriate compressive substrate ($S_m < 0$), and large strain magnitudes may be achieved when the critical thickness for the generation of misfit dislocations is not exceeded. Hence the depolarizing-field-induced increase of $a_3^{**}$ can be compensated down to the film thickness $t^* \cong c_{11} / [4(q_{11} c_{12} - q_{12} c_{11})(-S_m) c_i]$.

Using Eqs. (1) and (5), we can calculate the out-of-plane polarization and depolarizing field as a function of the film thickness at a constant misfit strain $S_m < 0$. In symmetric MFM heterostructures ($W_1 = W_2$), the spontaneous polarization $P_s = P_3(V_a = 0)$ is progressively suppressed in thinner films due to the depolarizing-field effect. For perovskite ferroelectrics, the coefficient $a_{33}^*$ involved in Eq. (1) is positive so that we may first evaluate $P_s(t)$ in the $P^4$ approximation. The calculation yields

$$P_s^2 = \frac{1}{2a_{33}^*}\left[\frac{\theta - T}{2\varepsilon_0 C} - 2\left(\frac{q_{11}c_{12}}{c_{11}} - q_{12}\right)S_m - \frac{1}{2(\varepsilon_0 + c_i t)}\right], \quad (7)$$

where $\theta$ and $C$ are the Curie-Weiss temperature and constant of the bulk ferroelectric crystal. Equation (7) shows that the strain effect shifts the polarization suppression to smaller film thicknesses. At a fixed temperature $T$, the out-of-plane polarization $P_s$ vanishes below a critical film thickness $t_0$, which was regarded as a size-induced phase transition [10,27]. From Eq. (7) it follows that $t_0$ is inversely proportional to the interfacial capacitance, being given by the formula

$$t_0 \cong \frac{1}{c_i}\left[\frac{\theta - T}{\varepsilon_0 C} - 4\left(\frac{q_{11}c_{12}}{c_{11}} - q_{12}\right)S_m\right]^{-1}. \quad (8)$$

It should be emphasized, however, that this expression for the thickness $t_0$ is valid only at temperatures $T > T_1 = \theta + 2\varepsilon_0 C\ (q_{11} + q_{12} - 2q_{12}c_{12}/c_{11})S_m$, where the ferroelectric $c$ phase transforms below $t_0$ into the paraelectric phase. The calculations show that at $T < T_1$ the depolarizing field leads to the *polarization rotation*, but not to the disappearance of ferroelectricity at $t < t_0$. Instead of the paraelectric phase, an in-plane polarization state ($P_1 \neq 0$, $P_2 \neq 0$, $P_3 = 0$) becomes stable in thinnest films, e.g., the *aa* phase in the single-domain case [15]. The critical thickness $t_0(T < T_1)$ is larger than $t_0$ given by Eq. (8) since the transition is of the first-order at $T < T_1$. This kind of thickness-induced phase transformation was not considered in the preceding publications [10,11,27], where only complete disappearance of ferroelectricity below $t_0$ was discussed.

At film thicknesses well above $t_0(T)$, the spontaneous polarization $P_s(t)$ should be calculated from Eq. (1) using the $P^6$ or higher-order approximation. We performed these calculations for $Pb(Zr_{0.5}Ti_{0.5})O_3$ and $BaTiO_3$ films grown on $SrTiO_3$, assuming the misfit strain $S_m$ to be equal to a thickness-independent value attained in fully strained MFM trilayers. For $Pb(Zr_{0.5}Ti_{0.5})O_3$ films, we employed the $P^6$ approximation with the material parameters [28] determined from the set used in Ref. 16. The polarization of $BaTiO_3$ films was calculated in the $P^8$ approximation using the thermodynamic parameters of $BaTiO_3$ obtained recently in Ref. [29] and the elastic and electrostrictive constants of this crystal [30]. Since the magnitude of the interfacial capacitance $c_i$ affects the spontaneous polarization only via the product $c_i t$, the dependences $P_s(t)$ corresponding to different electrode materials can be described by one universal curve $P_s(t_{\text{eff}})$. Here the effective film thickness $t_{\text{eff}}$ is directly proportional to the actual thickness $t$ and may be defined as $t_{\text{eff}} = (c_i/c_1)t$ , where $c_1 = 1\ \text{F/m}^2$.

The dependences $P_s(t_{\text{eff}})$ calculated for strained $Pb(Zr_{0.5}Ti_{0.5})O_3$ and $BaTiO_3$ films are shown in Fig. 2. It can be seen that just above $t_0$ the spontaneous polarization steeply increases with thickness and reaches values comparable to the bulk polarization. At $c_i = 0.444\ \text{F/m}^2$ characteristic of $SrRuO_3$ electrodes [13], for example, Eq. (8) gives $t_0 \cong 2$ nm for $Pb(Zr_{0.5}Ti_{0.5})O_3$ films and $t_0 \cong 2.6$ nm for $BaTiO_3$ films. Therefore, even nanoscale capacitors and tunnel junctions may have the out-of-plane polarization sufficient for the memory applications.

From Eqs. (1) and (5) we can also calculate the thickness dependence of the depolarizing field, which is described by a universal curve $E_{\text{dep}}(t_{\text{eff}})$. Figure 2 demonstrates that $E_{\text{dep}}$ varies nonmonotonically with the film thickness, reaching maximum at some thickness above $t_0$. Remarkably, the maximum value of $E_{\text{dep}}$ is independent of the interfacial capacitance $c_i$. Comparing this value with the electric field $E_g = \Delta_g/(et)$ necessary for the generation of free carriers inside a ferroelectric having the band gap $\Delta_g$ [6], we can check the validity of our assumption that ferroelectric behaves as an insulator. For $BaTiO_3$ and $Pb(Zr_{0.5}Ti_{0.5})O_3$, the band gap is about 3.5 eV so that in the discussed nanoscale range $E_g$ appears to be much larger than $E_{dep}$ [$E_g$ reduces below the maximum depolarizing field only at film thicknesses larger than 19 nm for $BaTiO_3$ and 9 nm for $Pb(Zr_{0.5}Ti_{0.5})O_3$].

### III. STABILITY OF SINGLE-DOMAIN STATE

Now we must investigate the stability of single-domain polarization state against the formation of a 180º domain pattern reducing the depolarizing field inside the film. To that end, it is necessary to analyze a wave-like perturbation of the uniform polarization state [31]. In contrast to the previous simplified treatments of similar stability problems [31,32,20], we present a rigorous formulation of the problem for perovskite ferroelectrics.

Since the polarization distribution becomes inhomogeneous, the equation of state (1) should be replaced by the Euler equations involving the gradient terms. For ferroelectrics with a cubic paraelectric phase, these equations in our two-dimensional case [33] become

$$\begin{aligned}&(2a_1 + 4a_{11}P_1^2 + 2a_{12}P_3^2 + \ldots)P_1 - 2q_{11}S_{11}P_1 \\ &- 2q_{12}(S_{22} + S_{33})P_1 - 2q_{44}S_{13}P_3 - E_1 \\ &= g_{11}P_{1,11} + g_{44}P_{1,33} + (g_{12} + g_{44})P_{3,13},\end{aligned} \quad (9)$$

$$\begin{aligned}&(2a_1 + 4a_{11}P_3^2 + 2a_{12}P_1^2 + \ldots)P_3 - 2q_{11}S_{33}P_3 \\ &- 2q_{12}(S_{11} + S_{22})P_3 - 2q_{44}S_{13}P_1 - E_3 \\ &= (g_{12} + g_{44})P_{1,13} + g_{44}P_{3,11} + g_{11}P_{3,33},\end{aligned} \quad (10)$$

where indices after the comma denote differentiation with respect to the coordinates $x_1$ and $x_3$, $g_{ln}$ are the coefficients of the gradient terms in the free-energy expansion [34], and $S_{ij}$ are the lattice strains in the film ($S_{22} = S_m$).

The polarization components involved in Eqs. (9)-(10) can be written as $P_3 = P_s + \delta P_3(x_3)\exp(ikx_1)$ and $P_1 = \delta P_1(x_3)\exp(ikx_1)$ with $\delta P_1$, $\delta P_3$ << $P_s$. Similar relations $S_{33} = (q_{11}P_s^2 - 2c_{12}S_m)/c_{11} + \delta S_{33}(x_3)\exp(ikx_1)$, $S_{11} = S_m + \delta S_{11}(x_3)\exp(ikx_1)$, $S_{13} = \delta S_{13}(x_3)\exp(ikx_1)$, and $\phi = \phi_0(x_3) + \delta\phi(x_3)\exp(ikx_1)$ can be introduced for the film strains and the electrostatic potential $\phi$. Using these formulae and retaining only the lowest-order perturbation terms, we obtain

$$(\tilde{\chi}_{11} + g_{11}k^2)\delta P_1 - 2q_{44}P_s\delta S_{13} + ik\delta\phi = g_{44}\delta P_{1,33} + ik(g_{12} + g_{44})\delta P_{3,3}, \quad (11)$$

$$(\tilde{\chi}_{33} + g_{44}k^2)\delta P_3 - 2q_{11}P_s\delta S_{33} - 2q_{12}P_s\delta S_{11} + \delta\phi_{,3} = ik(g_{12} + g_{44})\delta P_{1,3} + g_{11}\delta P_{3,33}, \quad (12)$$

where $\tilde{\chi}_{11} = 2a_1 - 2(q_{11} + q_{12} - 2q_{12}c_{12}/c_{11})S_m + 2(a_{12} - q_{11}q_{12}/c_{11})P_s^2 + 2a_{112}P_s^4 + ...$ and $\tilde{\chi}_{33} = 2a_1 + 4[q_{11}(c_{12}/c_{11}) - q_{12}]S_m + 12[a_{11} - q_{11}^2/(6c_{11})]P_s^2 + 30a_{111}P_s^4 + ...$. It should be noted that $\tilde{\chi}_{11}$ and $\tilde{\chi}_{33}$ depend on the interfacial capacitance $c_i$ and the film thickness $t$ because the spontaneous polarization $P_s$ is a function of these parameters.

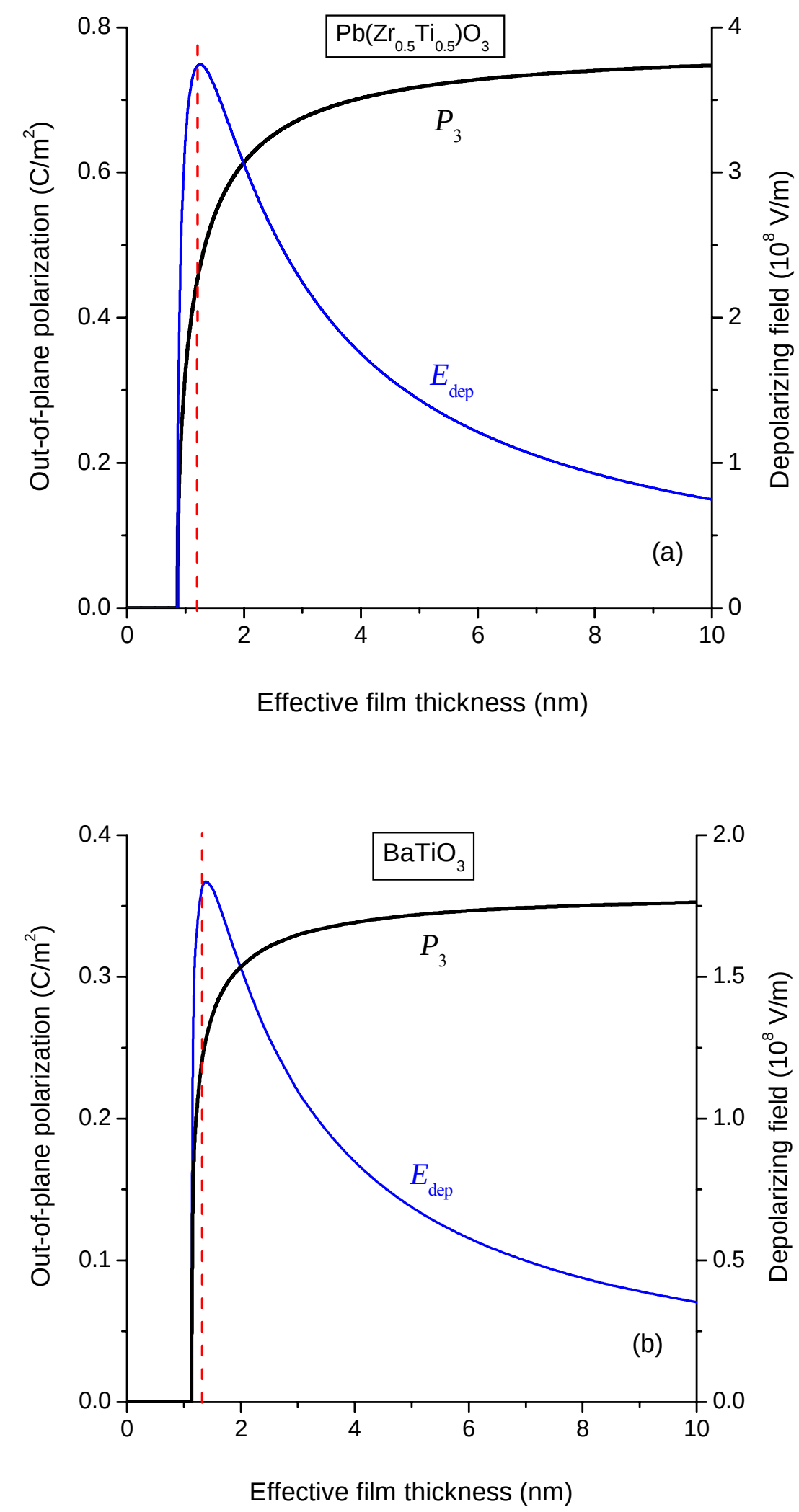

FIG. 2. Thickness dependence of the out-of-plane polarization and depolarizing field in ultrathin $Pb(Zr_{0.5}Ti_{0.5})O_3$ (a) and $BaTiO_3$ (b) films epitaxially grown on $SrTiO_3$. The misfit strain is taken to be $-26\times10^{-3}$ for $BaTiO_3$ films [17] and $-39\times10^{-3}$ for $Pb(Zr_{0.5}Ti_{0.5})O_3$ films [18]; the temperature equals 25 ºC. The actual film thickness $t$ corresponding to a given effective thickness $t_{eff}$ can be found from the relation $t = (c_1/c_i)\, t_{eff}$, where $c_1$ = 1 F/m$^2$ and $c_i$ is the total capacitance of the electrodes. The dashed line shows the film thickness below which the single-domain state becomes unstable.

In addition to Eqs. (11)-(12), several other relationships must be satisfied. First, the electrostatic condition div **D** = 0 holding inside an insulating film yields the following relation between the perturbation $\delta\phi$ of electrostatic potential and the polarization perturbations:

$$ik\delta P_1 + \delta P_{3,3} + \varepsilon_0(k^2\delta\phi - \delta\phi_{,33}) = 0\,. \quad (13)$$

Second, the strains $S_{ij}$ in the film must obey the classical compatibility condition $e_{ikl}e_{jmn}S_{ln,km} = 0$, where $e_{ikl}$ is the permutation symbol. In our case, this condition reduces to

$$\delta S_{11,33} - k^2\delta S_{33} = 2ik\delta S_{13,3}\,. \quad (14)$$

Finally, from the equations of mechanical equilibrium $\sigma_{ij,j} = 0$ written for the stresses $\sigma_{ij}$ in the film we obtain

$$ik(c_{11}\delta S_{11} + c_{12}\delta S_{33}) + 4c_{44}\delta S_{13,3} = 2q_{44}P_s\delta P_{1,3} + 2ikq_{12}P_s\delta P_3\,, \quad (15)$$

$$c_{12}\delta S_{11,3} + c_{11}\delta S_{33,3} + 4ik\,c_{44}\delta S_{13} = 2ikq_{44}P_s\delta P_1 + 2q_{11}P_s\delta P_{3,3}\,. \quad (16)$$

Equations (11)-(16) form a system of six differential equations for six unknown functions: $\delta P_1$, $\delta P_3$, $\delta\phi$, $\delta S_{11}$, $\delta S_{33}$, and $\delta S_{13}$. The analysis of this system shows that the terms involving very small factor $\varepsilon_0$ may be neglected in comparison with other terms. Then, after some mathematical manipulation, Eqs. (11)-(16) may be reduced to the following system of two simultaneous equations:

$$\left[\tilde{\chi}_{33} - 2q_{12}\frac{(2q_{11} - q_{44})}{(c_{12} + 2c_{44})}P_s^2 + g_{44}k^2\right]k^2\delta P_3 - \left[\tilde{\chi}_{11} + 2q_{44}\frac{(2q_{11} - q_{44})}{(c_{12} + 2c_{44})}P_s^2 + 2(g_{11} - g_{12} - g_{44})k^2\right]\delta P_{3,33} + g_{44}\delta P_{3,3333} - \left[2q_{12} - c_{11}\frac{(2q_{11} - q_{44})}{(c_{12} + 2c_{44})}\right]P_s k^2\delta S_{11} - \left[q_{44} + 2c_{44}\frac{(2q_{11} - q_{44})}{(c_{12} + 2c_{44})}\right]P_s\delta S_{11,33} = 0, \quad (17)$$

$$4c_{44}q_{12}P_s k^4\delta P_3 + 2[(c_{12} + 2c_{44})q_{11} - c_{11}q_{12} - c_{12}q_{44}] \times P_s k^2\delta P_{3,33} - 2c_{11}q_{44}P_s\delta P_{3,3333} - 2c_{11}c_{44}k^4\delta S_{11} + (c_{11}^2 - c_{12}^2 - 4c_{12}c_{44})k^2\delta S_{11,33} - 2c_{11}c_{44}\delta S_{11,3333} = 0. \quad (18)$$

It should be emphasized that Eqs. (17)-(18) differ dramatically from the differential equation used to describe the stability of a uniform polarization state formerly [31,20]. Indeed, owing to the electrostrictive coupling between the polarization and strain, Eq. (17) contains terms depending on the lattice strain and its second derivative, which were totally ignored previously. Moreover, additional Eq. (18) must be satisfied because the polarization wave creates an elastic wave in a piezoelectric medium.

Since the system (17)-(18) is homogeneous, we may seek the solution in the form of $\delta P_3 = A\exp(\lambda x_3)$ and $\delta S_{11} = B\exp(\lambda x_3)$. The substitution of these formulae into Eqs. (17)-(18) leads to a homogeneous system of two linear algebraic equations in the unknown coefficients $A$ and $B$. Calculating the determinant $\Delta_\lambda$ of this system and setting it to zero, we obtain the characteristic equation $\Delta_\lambda(\lambda) = 0$ being a quartic algebraic equation with respect to $\lambda^2$ in our case. When all roots $\lambda_n(k)$ of this equation are distinct, the sought perturbations can be written as $\delta P_3 = \sum_{n=1}^{8} A_n\exp(\lambda_n x_3)$ and $\delta S_{11} = \sum_{n=1}^{8} B_n\exp(\lambda_n x_3)$. Here each coefficient $B_n$ can be expressed in terms of $A_n$ using Eq. (18), for example. The calculation yields

$$B_n = 2P_s A_n \times \frac{2c_{44}q_{12}k^4 + [(c_{12}+2c_{44})q_{11} - c_{11}q_{12} - c_{12}q_{44}]k^2\lambda_n^2 - c_{11}q_{44}\lambda_n^4}{2c_{11}c_{44}k^4 - (c_{11}^2 - c_{12}^2 - 4c_{12}c_{44})k^2\lambda_n^2 + 2c_{11}c_{44}\lambda_n^4} \quad (19)$$

In turn, the coefficients $A_n$ must satisfy a system of eight simultaneous equations, which follow from the boundary conditions on the problem.

To derive this system of equations, we first employ the continuity of potential $\phi$ and electric displacement $D_3$ at the film-electrode interfaces. Within the film, the variation $\delta\phi(x_3)$ of electrostatic potential can be calculated in terms of $A_n$ using Eqs. (11)-(16). The potential $\phi_m$ inside the electrodes in the presence of polarization wave becomes $\phi_m = \phi_{m0}(x_3) + \delta\phi_m(x_3)\exp(ikx_1)$. The perturbation $\delta\phi_m(x_3)$ can be easily found in the screening length approximation to be $\delta\phi_m = L_2\exp(-\zeta x_3)$ with $\zeta = \sqrt{k^2 + l_s^{-2}}$ in the right electrode ($x_3 \geq 0$) and $\delta\phi_m = L_1\exp[\zeta(x_3 + t)]$ in the left one ($x_3 \leq -t$). Using these expressions to formulate the boundary conditions and eliminating then the constants $L_1$ and $L_2$, we obtain $\delta P_3(x_3 = 0) = \varepsilon_0\varepsilon_m\zeta\ \delta\phi(x_3 = 0)$ and $\delta P_3(x_3 = -t) = -\varepsilon_0\varepsilon_m\zeta\ \delta\phi(x_3 = -t)$. These relationships give us the first two equations for $A_n$. Another two equations follow from the conditions imposed on the polarization derivative $P_{3,3}$ [14], which in the approximation of an infinite extrapolation length reduce to $\delta P_{3,3} = 0$ at $x_3 = 0$ and $x_3 = -t$.

Proceed now to the mechanical boundary conditions imposed on an epitaxial thin film grown on a nonpiezoelectric cubic substrate. Assuming the bottom electrode to be fully strained by a thick substrate, we shall neglect the mechanical influence of electrodes in comparison with the substrate effect. Then the conditions on the top film surface reduce to the absence of stresses $\sigma_{33}$ and $\sigma_{13}$ at $x_3 = 0$. The lattice matching on the bottom film surface implies the continuity of the mechanical displacement and the stresses $\sigma_{33}$ and $\sigma_{13}$ at $x_3 = -t$. It can be shown that the first of these conditions in our case gives $\delta S_{11} = \delta S_{11}^{sub}$ and $2\delta S_{13} - (1/ik)\delta S_{11,3} = 2\delta S_{13}^{sub} - (1/ik)\delta S_{11,3}^{sub}$ at $x_3 = -t$. The strains $S_{ij}^{sub} = \delta S_{ij}^{sub}(x_3)\exp(ikx_1)$ and the stresses $\sigma_{ij}^{sub} = \delta\sigma_{ij}^{sub}(x_3)\exp(ikx_1)$ in the substrate can be calculated from the compatibility condition and the equations of mechanical equilibrium. For the nonzero strains, the calculation gives

$$\delta S_{11}^{sub} = R_1\exp(\beta_1 kx_3) + R_2\exp(\beta_2 kx_3), \quad (20)$$

$$\delta S_{33}^{sub} = -\frac{(c_{11} - 2c_{44}\beta_1^2)}{(c_{12} + 2c_{44})} R_1\exp(\beta_1 kx_3) - \frac{(c_{11} - 2c_{44}\beta_2^2)}{(c_{12} + 2c_{44})} R_2\exp(\beta_2 kx_3), \quad (21)$$

$$\delta S_{13}^{sub} = \frac{1}{4ic_{44}}\left\{\left[\frac{c_{11}(c_{11} - 2c_{44}\beta_1^2)}{c_{12} + 2c_{44}} - c_{12}\right]\beta_1 R_1\exp(\beta_1 kx_3) + \left[\frac{c_{11}(c_{11} - 2c_{44}\beta_2^2)}{c_{12} + 2c_{44}} - c_{12}\right]\beta_2 R_2\exp(\beta_2 kx_3)\right\}, \quad (22)$$

where $\beta_{1,2} = \sqrt{\xi \pm \sqrt{\xi^2 - 1}}$, $\xi = (c_{11}^2 - c_{12}^2 - 4c_{12}c_{44})/(4c_{11}c_{44})$ and $c_{mn}$ are the elastic stiffnesses of the substrate assumed to be equal to the film stiffnesses for simplicity. Similar relations can be derived for the functions $\delta\sigma_{ij}^{sub}(x_3)$ defining the stresses in the substrate. The coefficients $R_1$ and $R_2$ represent additional unknown quantities which can be calculated in terms of $A_n$ using the strain conditions on the film/substrate interface. The remaining four stress conditions together with the four electrical conditions give us the sought system of eight linear algebraic equations for the constants $A_n$.

Since this system is homogeneous, a nonzero solution for the set of coefficients $A_n$ exists only if the determinant $\Delta_A$ of this system equals zero. By solving the equation $\Delta_A(k) = 0$ numerically, it is possible to check the existence of any root $k \neq 0$ at a given film thickness $t$, interfacial capacitance $c_i$, temperature $T$, and misfit strain $S_m$. If there are no such roots, the uniform polarization state remains stable against inhomogeneous polarization perturbations.

The critical thickness $t_c$, at which the single-domain state becomes unstable, can be found as a maximum value of $t$ at which a nonzero solution for $k$ first appears.

We performed necessary numerical calculations for fully strained $BaTiO_3$ and $Pb(Zr_{0.5}Ti_{0.5})O_3$ films grown on $SrTiO_3$. Since for our purposes it is sufficient to determine the upper bound $t_{c,\max}$ of the critical thickness, we simplified the problem by setting the gradient coefficients $g_{ln}$ to zero. (Note that the polarization gradient terms can only decrease the critical thickness and that the values of $g_{ln}$ are not exactly known for perovskite ferroelectrics). Then the characteristic equation $\Delta_\lambda(\lambda) = 0$ reduces to a cubic equation with respect to $\lambda^2$, and the roots $\lambda_n$ of this equation become proportional to the wave number $k$.

Since the critical thickness is inversely proportional to the interfacial capacitance $c_i$, it is sufficient to compute $t_{c,\max}$ for one particular value of $c_i$ only. We studied the case of capacitors with two $SrRuO_3$ electrodes assuming $c_i = 0.444$ F/m$^2$ [35]. At room temperature, $t_{c,\max}$ was found to be slightly below 2.98 nm for $BaTiO_3$ films and below 2.73 nm for $Pb(Zr_{0.5}Ti_{0.5})O_3$ films. The magnitude of $t_{c,\max}$ weakly decreases with decreasing temperature, reducing down to about 2.46 nm in $BaTiO_3$ films and to about 2.63 nm in $Pb(Zr_{0.5}Ti_{0.5})O_3$ films at $T = -200$ °C. The dependence of $t_{c,\max}$ on the interfacial capacitance $c_i$ at $T = 25$ °C is shown in Fig. 3.

Our numerical calculations also demonstrated that the critical thickness $t_c$ is always smaller than the thickness $t_{th}$ at which the film inverse susceptibility $\chi_{33} = 2a_3^* + 12a_{33}^* P_s^2 + 30a_{111} P_s^4 + \ldots$ goes to zero. Since the "threshold" thickness $t_{th}$ becomes very close to $t_c$ when the latter corresponds to the nanoscale range [at room temperature, $t_{th} \cong 3.1$ nm for $BaTiO_3$ films and $t_{th} \cong 2.81$ nm for $Pb(Zr_{0.5}Ti_{0.5})O_3$ films], the equation $\chi_{33}(t) = 0$ may be used to roughly estimate the upper bound of $t_c$ without performing complicated numerical calculations.

Thus, by combining highly strained epitaxial films with metallic electrodes having good screening properties it is possible to stabilize the single-domain ferroelectric state in nanoscale capacitors and tunnel junctions [36].

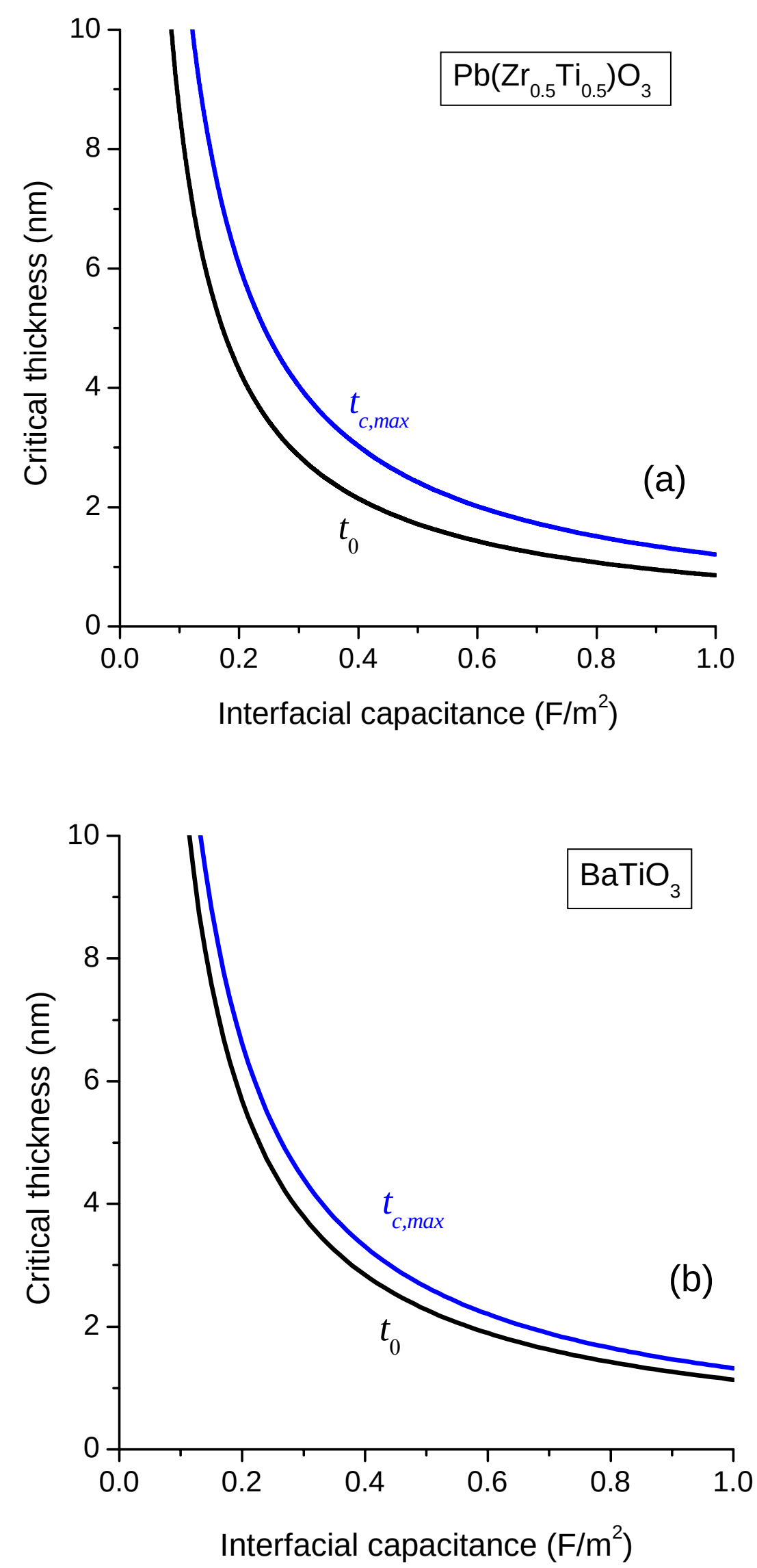

FIG. 3. Critical thicknesses of strained $Pb(Zr_{0.5}Ti_{0.5})O_3$ (a) and $BaTiO_3$ (b) films as a function of the total interfacial capacitance characterizing the screening space charge in the electrodes. Here $t_0$ is the thickness at which the out-of-plane polarization goes to zero, and $t_{c,\max}$ is the upper bound of the critical thickness below which the single-domain $c$ phase becomes unstable with respect to the formation of 180º domains.

Remarkably, this stabilization results from the elastic effect caused by the electrostrictive coupling between polarization and strain, which was totally ignored previously [31,20].

## IV. DEPOLARIZING-FIELD EFFECT ON ELECTRON TUNNELING

Proceed now to the asymmetric MFM heterostructures that involve dissimilar electrodes. When the electrode work functions $W_1$ and $W_2$ are different, a nonzero electric field $E_3 \cong (W_2 - W_1)/(et)$ acts on the film in a short-circuited capacitor [25]. This field increases with decreasing film thickness and becomes very strong in ultrathin films. In the case of one $SrRuO_3$ and one Pt electrode, for example, we have $W_1 \approx 5.2$ eV [37] and $W_2 \approx 5.65$ eV [38] so that the field intensity exceeds $10^8$ V/m at $t < 4.5$ nm. Evidently, the film becomes homogeneously polarized under such field, with the polarization pointing to the electrode having higher work function. Since this polarization is only weakly dependent on temperature, the ferroelectric to paraelectric transition disappears in a film sandwiched between short-circuited dissimilar electrodes.

The usual temperature dependence of polarization, however, restores when a bias voltage $V_a = (W_1 - W_2)/e$ is applied to the electrodes. Hence the plots shown in Fig. 2 now describe the thickness dependence of the polarization $P_3(V_a)$ appearing at this voltage. The polarization reversal also occurs in asymmetric capacitors, but the hysteresis loop becomes shifted along the voltage axis.

Another specific feature of asymmetric MFM heterostructures is associated with the screening of polarization charges by dissimilar electrodes. When the screening abilities of two electrodes are different, the depolarizing field changes the *mean value* of electrostatic potential in the heterostructure [3,4]. This feature is expected to play an important role in ferroelectric tunnel junctions (FTJs), where the quantum mechanical electron tunneling represents the dominant conduction mechanism. Indeed, since the sign of the above effect depends on the polarization orientation in the ferroelectric barrier, the polarization reversal occurring at the coercive voltage $V_c$ should be accompanied by a change of the barrier conductance. Such a resistive switching of ferroelectric origin was studied theoretically in Refs. [3,4], but a linear approximation was employed for the field dependence of polarization.

The nonlinear theory enables us to calculate the depolarizing-field-induced change $\Delta\bar{\phi}$ of the mean potential $\bar{\phi}$ in the ferroelectric barrier rigorously. Since in our case $\bar{\phi}$ is equal to the arithmetic mean of electrostatic potentials on the film surfaces, $\Delta\bar{\phi}$ is defined by the semi-difference of the voltage drops $\Delta\phi_{m1}$ and $\Delta\phi_{m2}$ in the electrodes. Using the explicit expressions for these quantities and Eq. (5), we find

$$\Delta\bar{\phi} = \frac{c_i\, t}{2(\varepsilon_0 + c_i t)} P_3\left(\frac{1}{c_{m2}} - \frac{1}{c_{m1}}\right) \cong \frac{1}{2} P_3\left(\frac{1}{c_{m2}} - \frac{1}{c_{m1}}\right), \qquad (23)$$

where $P_3 = P_3(t)$ is the thickness-dependent equilibrium polarization in the barrier. Equation (23) demonstrates that the depolarizing-field effect on the mean potential is directly proportional to the difference between the reciprocal capacitances of two electrodes. The magnitude of $\Delta\bar{\phi}$ is expected to be large when one electrode is made of a conducting perovskite material like $SrRuO_3$, whereas the other electrode is produced of an elemental metal like Pt. Accordingly, taking $c_{m1} \approx 0.9$ F/m$^2$ [13], $c_{m2} \approx 0.4$ F/m$^2$ [24], and estimating $P_3$ from Fig. 2, we find that at $V_a = (W_1 - W_2)/e$ the change $\Delta\bar{\phi}$ can be as large as 0.2 V in $BaTiO_3$ films and 0.4 V in $Pb(Zr_{0.5}Ti_{0.5})O_3$ films.

The depolarizing-field effect on the conductance of a FTJ can be evaluated in the approximation of an average barrier [39,3]. Depending on the polarization orientation in the junction, $\Delta\bar{\phi}$ either increases or reduces the mean barrier height, which leads to the appearance of two different resistance states. When the bias voltage $V_a = (W_1 - W_2)/e$ is small, the ratio of conductances $G_L$ and $G_H$, which characterize the low- and high-resistance states at $V = V_a$, can be estimated in the linear approximation as

$$\frac{G_L}{G_H} \cong \frac{\left(1 + \frac{t}{t_\varphi}\sqrt{1 - \frac{e|\Delta\bar{\phi}|}{\varphi_0}}\right)\exp\left(-\frac{t}{t_\varphi}\sqrt{1 - \frac{e|\Delta\bar{\phi}|}{\varphi_0}}\right)}{\left(1 + \frac{t}{t_\varphi}\sqrt{1 + \frac{e|\Delta\bar{\phi}|}{\varphi_0}}\right)\exp\left(-\frac{t}{t_\varphi}\sqrt{1 + \frac{e|\Delta\bar{\phi}|}{\varphi_0}}\right)}, \qquad (24)$$

where $t_\varphi = h/(4\pi\sqrt{2m_f^*\varphi_0})$, $\varphi_0$ is the average barrier height at zero polarization, and $m_f^*$ is the electron effective mass in the barrier, which may differ considerably from the free electron mass $m_e$ [40,41]. At $\varphi_0$ = 0.5 eV, $|\Delta\bar{\phi}|$ = 0.1 V, and $t$ = 3.2 nm, for instance, Eq. (24) gives $G_L/G_H$ about 7 for $m_f^* = 0.2m_e$ and about 600 for $m_f^* = 2m_e$. At large bias voltages $V_a$, the quadratic and cubic terms in the current-voltage dependence should be also taken into account for the correct evaluation of $G_L$ and $G_H$ [3]. For FTJs involving Pt and $SrRuO_3$ electrodes, e.g., the ratio $G_L/G_H$ at the bias voltage $V_a \approx 0.45$ V appears to be about 9 for $m_f^* = 0.2m_e$ and ~ 900 for $m_f^* = 2m_e$. Thus, the conductance on/off ratio rises steeply with the increase of effective mass.

## V. CONCLUSION

In this paper, we have shown that the single-domain polarization state may be stabilized in ultrathin ferroelectric films by compressive in-plane strains counteracting the depolarizing-field effect. The reduction of spontaneous polarization, which is caused by $E_{dep}$, depends on the total capacitance of the screening charge in the electrodes (but not on their screening lengths as supposed in some recent publications [4,10,12]). In contrast, the depolarizing-field effect on the tunnel current through an ultrathin ferroelectric barrier is predominantly governed by the difference between the reciprocal capacitances of two electrodes.

## ACKNOWLEDGMENT

The financial support of the Deutsche Forschungsgemeinschaft is gratefully acknowledged. The research described in this paper was also supported by the Volkswagen-Stiftung project "Nano-sized ferroelectric hybrids" under contract No I/77737.